\documentclass[aps,reprint,showpacs,floatfix]{revtex4-2}

\usepackage{graphicx}
\usepackage{amssymb}
\usepackage{amsmath}
\usepackage{subfig}
\usepackage{caption}
\usepackage{hyperref}

\def\bea{\begin{eqnarray}}
\def\eea{\end{eqnarray}}
\def\beq{\begin{equation}}
\def\eeq{\end{equation}}

\def\bm{\begin{math}}
\def\me{\end{math}}

\begin{document}
\title{Universal Aging Dynamics and Scaling Laws in Three-Dimensional Driven Granular Gases}
\author{Rameez Farooq Shah$^{1}$, Syed Rashid Ahmad$^{1}$} 
\email{rmzshah@gmail.com,\\
srahmad@jmi.ac.in}
\affiliation{
$^{1}$Department of Physics, Jamia Millia Islamia, New Delhi 110025, India
}

\begin{abstract}
We establish universal scaling laws and quantify aging in three-dimensional uniformly heated hard sphere granular gases through large-scale event-driven molecular dynamics ($N=500{,}000$). We report three primary quantitative discoveries: (i) The characteristic energy decay time exhibits a universal inverse scaling $\tau_0 \propto \epsilon^{-1.03 \pm 0.02}$ with the dissipation parameter $\epsilon = 1 - e^2$. (ii) The steady-state temperature follows a precise power-law $T_{\mathrm{steady}} \propto \epsilon^{-1.51 \pm 0.03}$, reflecting the non-linear balance between thermostat heating and collisional dissipation. (iii) The velocity autocorrelation function $\bar{A}(\tau_w, \tau)$ demonstrates pronounced aging, with decay rates $\lambda$ following a power-law slowing down $\lambda(\tau_w) \propto \tau_w^{-0.82 \pm 0.05}$. These results establish the first 3D quantitative benchmarks for aging in driven dissipative gases, where near-Gaussian statistics persist despite extreme structural clustering.
\end{abstract}

\pacs{45.70.-n, 05.20.Dd, 05.70.Ln, 47.70.Nd}
\keywords{Granular gas, aging, velocity autocorrelation, inelastic collisions, white noise thermostat, non-equilibrium steady state}
\maketitle

\section{\label{introd} INTRODUCTION}
Granular materials, composed of macroscopic particles that interact via dissipative collisions, are ubiquitous in nature and find extensive applications in diverse fields ranging from civil engineering and pharmaceutical processing to astrophysics \cite{rmp_behringer, rmp_kadanoff, rmp_tsimring, duran, ristow}. When such systems are sufficiently fluidized, they exhibit behaviors reminiscent of molecular gases, albeit with fundamental differences stemming from the continuous dissipation of kinetic energy. The study of these ``granular gases'' has emerged as a vibrant frontier in non-equilibrium statistical mechanics, providing a rich testing ground for kinetic theory and hydrodynamic descriptions in systems far from thermal equilibrium \cite{nb_ktgg, cc70, haff83}.

A defining characteristic of granular gases is the spontaneous emergence of spatial structures, most notably in the form of high-density clusters \cite{gz93, mcny9296, sl9899}. This clustering arises as a direct consequence of inelastic collisions, which tend to reduce the relative velocity between particles and lead to local alignment. In the absence of an external energy source, the system undergoes homogeneous cooling, governed by Haff's law \cite{haff83}. However, this cooling process is often interrupted by the development of instabilities and the formation of these dense structures, which further accelerate energy dissipation and drive the system toward highly non-equilibrium states characterized by non-Gaussian velocity distributions \cite{vne98, bpepl2006, gs95}.

To maintain a steady state, energy must be continuously supplied to the system, typically through various thermostatting mechanisms \cite{willmac96, william96}. A common approach is the use of a white-noise thermostat, which mimics the effect of vibrating boundaries or external forcing by adding random velocity increments to all particles \cite{vne98, sl9899, ap0607}. Such driven systems settle into a non-equilibrium steady state where energy injection balances collisional dissipation, allowing for the investigation of long-time relaxation dynamics and transport properties \cite{tpcvn9798, jjbrey9698, adsp1213}.

A particularly intriguing aspect of non-equilibrium dynamics is the concept of aging and memory effects. Aging refers to the dependence of a system's relaxation properties on its age or waiting time \cite{bnk02, pdsp2018}. While granular systems are inherently dissipative, the study of aging in granular gases has predominantly focused on the freely cooling regime, where the number of collisions grows logarithmically with time \cite{bnk02, pdsp2018}. The extension of these concepts to driven, steady-state systems remains a relatively unexplored area, especially in three-dimensional (3D) geometries.

The novelty and significance of this work lie in its systematic quantification of aging in the 3D driven regime. By employing large-scale molecular dynamics simulations ($N=500{,}000$), we provide the first rigorous benchmarks for the slowing-down of velocity correlations in uniformly heated gases. Our analysis (i) identifies the precise power-law scaling governing the steady-state temperature $T_{\mathrm{steady}} \propto \epsilon^{-1.51}$, (ii) reveals a universal power-law decay of aging kinetics $\lambda(\tau_w) \propto \tau_w^{-0.82}$, and (iii) demonstrates that statistical near-Gaussianity persists even as structural clustering becomes extreme. These results bridge the gap between microscopic collisional dynamics and macroscopic relaxation in dissipative gases.

The paper is organized as follows. In Sec.\ref{themodel}, we provide a detailed description of our model for a uniformly heated hard sphere granular gas and the simulation protocol. In Sec.\ref{vdf}, we define the velocity autocorrelation function and discuss the theoretical framework for aging. In Sec.\ref{sim_details}, we describe our simulation protocol and parameter space. In Sec.\ref{results}, we present comprehensive results from our simulations, emphasizing the scaling laws and velocity statistics. Finally, Sec.\ref{sec:summry} summarizes our findings and explores their implications for the broader study of driven granular systems.

%\newpage
\section{\label{themodel} Formalism}
Our starting point is a homogeneous granular gas in three dimensions ($d=3$), consisting of identical spherical molecules. Without loss of generality, we may assume mass and diameters of the molecules to be unity. For hard sphere molecules, the post-collision velocities of the particles labeled as $i$ and $j$ are given as a function of pre-collision velocities by the rule:
\bea
\label{coll}
	 \vec{v}_i^{\prime} = \vec{v}_i - \frac{1+e}{2} \left[ \hat{n} \cdot \left( \vec{v}_i - \vec{v}_j \right) \right] \hat{n}, \nonumber \\
	 \vec{v}_j^{\prime} = \vec{v}_j + \frac{1+e}{2} \left[ \hat{n} \cdot \left( \vec{v}_i - \vec{v}_j \right) \right] \hat{n},
\eea
where $e(<1)$ is the coefficient of restitution. Here, $\hat{n}$ is a unit vector pointing from the position of particle $j$ to the position of particle $i$.

Much like the molecular gas, we  can associate a temperature with the granular gas. This temperature, called the granular temperature, is defined as $T=\left\langle\vec{v}^2\right\rangle / d$, where $\left\langle\vec{v}^2\right\rangle$ is the mean-squared velocity, and $d$ is the dimensionality. In the early stages and in the absence of any external drive, the time rate of change of granular temperature is given by \cite{haff83}
\begin{equation}
\label{cool}
\frac{dT}{dt}=-\frac{\epsilon\omega(T)T}{d}, \quad \epsilon=1-e^2 ,
\end{equation}
where $\omega(T)$ represents the frequency of collision at temperature $T$. From kinetic theory of gases, we know that $\omega(T)$ is given by  \cite{cc70}:
\begin{equation}
\label{omega}
\omega(T)\simeq \frac{2\pi^ {(d-1)/2} }{\Gamma (d/2)}~\chi(n)nT^{1/2} ,
\end{equation}
where $\chi(n)$ represents pair correlation function at contact for hard spheres with density $n$. Using equations~(\ref{cool}) and (\ref{omega}), we arrive at the Haff's law for the HCS:
\begin{equation}
\label{eq:haff_t}
T(t)=T_0 \left[ 1 + \frac{\epsilon\omega(T_0)}{2d}~t\right]^{-2} ,
\end{equation}
where $T_0$ is the initial temperature. 
If we define the average number of collisions in time $t$ as $\tau $, then  
\bea
\label{eq:tau_t}
\tau(t) &=& \int^{t}_{0}dt'\omega(t') \nonumber \\
&=& \frac{2d}{\epsilon} \ln \left[1 + \frac{\epsilon\omega(T_0)}{2d}t\right] .
\eea
As the system loses energy, the number of collisions increases logarithmically (instead of a linear increase) with time. The Haff's law as a function of $\tau$, can be written in the following form:  
\begin{equation}
\label{haff}
T(\tau)=T_0 \exp \left(-\frac{2\epsilon}{d}~\tau \right).
\end{equation}
In our 3D simulations ($d=3$), this corresponds to a decay factor of $2\epsilon/3$.

In presence of external driving, the injected energy compensates for the loss due to collisions, and the system settles to a non-equilibrium steady state. For a driven granular system, the stochastic equation of motion is described as:
\begin{equation}
\label{eq:motion}
    \dfrac{d \vec{v}_i}{dt} = \dfrac{\vec{F}_i^c}{m} + \dfrac{\vec{F}_i^t}{m}, 
\end{equation}
where $m$ is the mass of the particle, $\vec{F}_i^c$ is the force on the $i^{\text{th}}$ particle due to pairwise collisions given by relation~\eqref{coll}, and $\vec{F}_i^t$ is the external stochastic driving force.
 The thermostat force is modeled as Gaussian white noise with zero mean and is uncorrelated for different particles and different Cartesian components:
\begin{equation}
    \langle F_{i,\alpha}^t (t) F_{j,\beta}^t (t')\rangle = \xi_0^2 \delta_{ij} \delta_{\alpha \beta} \delta(t-t'), 
\end{equation}
\begin{equation}
\label{eq:mean_force}
    \langle \vec{F}_i^t(t) \rangle = 0,
\end{equation}
where $\alpha, \beta \in \{x,y,z\}$, $\xi_0$ characterizes the strength of the stochastic force, which relates to the diffusion in velocity space. We set $\xi_0^2 = 2r T(0)$ to ensure initial compatibility with the thermal state. With $T(0) = 1$ in reduced units, this yields $\xi_0^2 = 0.002$. This white noise thermostat uniformly heats all particles and drives the system to a non-equilibrium steady state.

\textit{Simulation Parameters}: We perform large-scale simulations with $N=500{,}000$ hard spheres in a cubic box with periodic boundary conditions. The number density is set to $n=0.02$, corresponding to a volume fraction $\phi = (n \pi \sigma^3)/6 \approx 0.01$, well within the dilute regime where binary collisions dominate. We utilize an event-driven molecular dynamics algorithm with a time-step $dt=0.01$ for the thermostat integration between collisions. To ensure statistical rigor, all results are averaged over 50 independent runs, and error bars throughout denote the standard error of the mean.
	
\section{\label{vdf} VELOCITY AUTOCORRELATION FUNCTION AND AGING}
The velocity autocorrelation function quantifies the persistence of velocity fluctuations and is defined as:
\begin{equation}
\label{eq:vacf}
A(\tau_w, \tau) = \frac{1}{N} \sum_{i=1}^{N} \langle \vec{v}_i(\tau_w) \cdot \vec{v}_i(\tau_w + \tau) \rangle,
\end{equation}
where $\tau_w$ is the waiting time (measured from the initial state at $\tau=0$), and $\tau$ is the correlation time. The angular brackets denote an ensemble average over independent realizations. We also study the normalized autocorrelation function:
\begin{equation}
\label{eq:norm_vacf}
\bar{A}(\tau_w, \tau) = \frac{A(\tau_w, \tau)}{A(\tau_w, 0)} = \frac{A(\tau_w, \tau)}{dT(\tau_w)},
\end{equation}
where the denominator is simply the granular temperature at time $\tau_w$ multiplied by the dimensionality.

The velocity autocorrelation function serves as a probe of the history and memory of the system's dynamics. During each collision, local parallelization of particle velocities occurs, leading to correlated motion and the formation of clusters---regions with higher and lower particle densities. The persistence of these velocity correlations is quantified by $A(\tau_w, \tau)$.

For systems in thermal equilibrium, the autocorrelation function depends only on the time difference $\tau$ due to time-translation invariance. However, for non-equilibrium systems, $A(\tau_w, \tau)$ depends independently on both $\tau_w$ and $\tau$, not merely on their difference. This independent dependence on the waiting time $\tau_w$ is the hallmark of aging.

For freely cooling granular gases (HCS), Ben-Naim and Krapivsky \cite{bnk02} and more recently Das et al. \cite{pdsp2018} have investigated the aging properties. It has been derived that the velocity autocorrelation decays as:
\begin{equation}
\label{eq:bnk1}
\bar{A}(\tau_w, \tau) = \exp \left( -\frac{d+2}{2d}(1 + \epsilon) \tau \right),
\end{equation}
or equivalently:
\begin{equation}
\label{eq:bnk2}
A(\tau_w, \tau) = dT(0) \exp \left( -\frac{2\epsilon}{d} \tau_w \right) \exp \left( -\frac{d+2}{2d}(1 + \epsilon) \tau \right).
\end{equation}
These expressions reveal the explicit $\tau_w$ dependence through the prefactor, demonstrating aging in the freely cooling case.

For the completely elastic case ($\epsilon = 0$, or $e=1$), thermal velocities decorrelate rapidly, and $A(\tau_w, \tau)$ decays to zero within a few collisions per particle. In contrast, for inelastic systems, dissipation leads to velocity alignment and clustering, resulting in longer-lived correlations. The strength of dissipation, controlled by the restitution coefficient $e$, thus plays a crucial role in determining the aging characteristics of the system.

\section{\label{sim_details} SIMULATION DETAILS}
\begin{figure}[!ht]
    \centering
    \includegraphics[width=1.0\columnwidth]{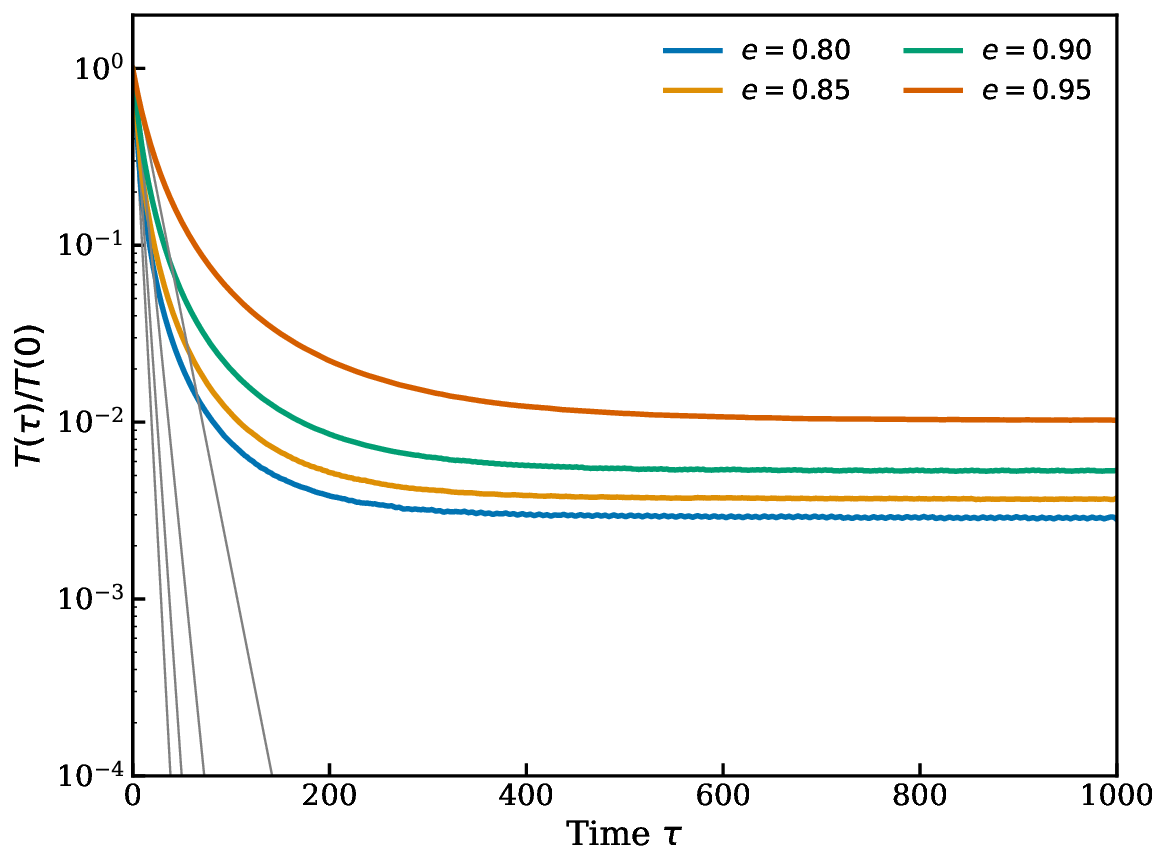} 
    \caption{Time dependence of the granular temperature in $d = 3$, shown on a semilog scale. We plot the normalized granular temperature $T(\tau)/T(0)$ vs $\tau$ for $e = 0.80, 0.85, 0.90$, and $0.95$. The solid lines denote Haff's law.}
    \label{fig_haff1}
\end{figure}

We perform event-driven molecular dynamics simulations of a system of hard sphere inelastic particles \cite{allentild, rapaport}. The system is initialized by assigning each particle a random position and velocity. Our system consists of $N=500{,}000$ particles confined in a 3D cubical box with periodic boundary conditions. The number density is set to $n=0.02$, which ensures the system remains in the dilute regime where binary collisions dominate and higher-order collisions are negligible. This large particle number is necessary to obtain statistically reliable results and to properly capture the collective clustering behavior characteristic of granular gases. The initial positions are generated such that no two particle cores overlap. The random velocity components are chosen from a Gaussian distribution with $\sum_i \vec{v}_i = 0$ to ensure zero total momentum.

All particles are identical with unit mass $m=1$ and diameter $\sigma=1$. To establish a well-defined initial condition, the system is first evolved to $\tau = 100$ at $e = 1$ (elastic collisions) without any energy input. This relaxation ensures that the velocity distribution reaches the Maxwell-Boltzmann form, eliminating any artifacts from the initial random configuration. This equilibrated state serves as the initial condition for our subsequent inelastic simulations.

The postcollision velocities are obtained from precollision velocities by Eq.~\eqref{coll}. The system is subject to Gaussian white noise heating, where a stochastic velocity increment is added to each particle after every time step $dt$ as follows:
\begin{equation}
\label{eq:noise}
    \vec{v}_i(t+dt) = \vec{v}_i(t) + \sqrt{r} \sqrt{dt}\, \xi,
\end{equation}
where $\xi$ is a random variable uniformly distributed in the interval $[-1/2, 1/2]$, and $r$ is the amplitude of the noise, chosen to be $r=0.001$. This value is selected to provide sufficient energy input to balance dissipation and reach a quasi-steady state, without overwhelming the collision dynamics or driving the system far from the dilute regime. After adjusting the velocities, the system is shifted to the center-of-mass frame to ensure conservation of linear momentum:
\begin{equation}
    \vec{v}_i = \vec{v}_i - \dfrac{1}{N} \sum_{i=1}^{N} \vec{v}_i. 
\end{equation}

We evolve the system until $\tau = 1000$ for four different values of the restitution coefficient: $e = 0.95, 0.90, 0.85,$ and $0.80$. All results presented here correspond to averages over $50$ independent initial conditions, providing robust statistics and allowing us to accurately compute ensemble averages.

\section{\label{results} RESULTS}
\subsection{Temperature Evolution and Steady State}

At $t = 0$, the system is initialized with a homogeneous density field and velocity components satisfying the Maxwell-Boltzmann distribution. Due to dissipative collisions, particles lose their kinetic energy, leading to a decay of the granular temperature. The thermostat noise partially compensates for the energy loss. This process continues until a non-equilibrium steady state is reached where the energy dissipated per collision balances the energy injected by the thermostat.

In Fig.~\ref{fig_haff1}, we plot the time evolution of the normalized temperature $T(\tau)/T(0)$ as a function of $\tau$ on a semilog scale for different values of the restitution coefficient $e$. For reference, Haff's law [Eq.~\eqref{haff}] is shown as solid lines. At early times, the system follows Haff's law closely, indicating homogeneous cooling. However, as the system evolves and the thermostat begins to balance dissipation, deviations from Haff's law become apparent, and the temperature approaches a steady-state value. The steady-state is reached earlier for weaker dissipation (larger $e$), as less energy input is required to balance the smaller energy loss per collision.

\begin{figure}[!ht]
    \centering
	\includegraphics[width=1.0\columnwidth]{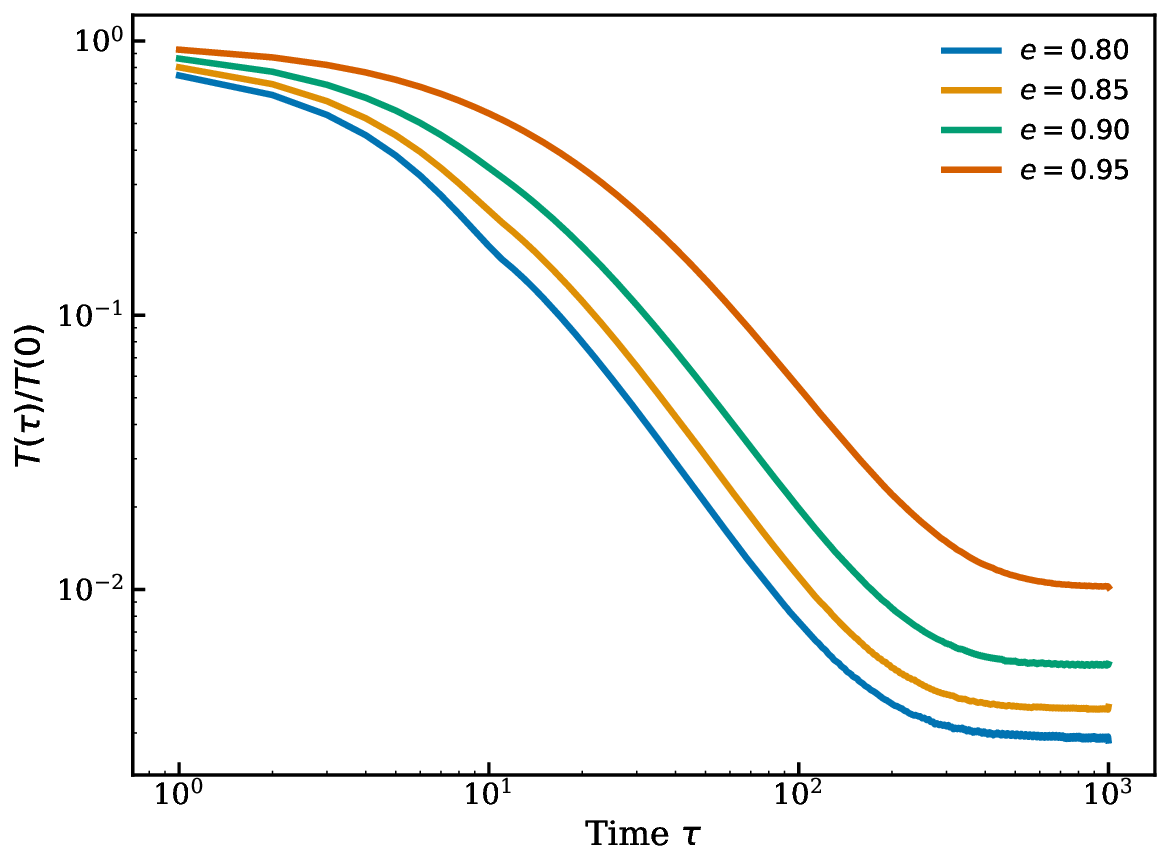} 
    \caption{Time dependence of the granular temperature in $d = 3$, shown on a log-log scale. It can be clearly seen that the temperature for each value of $e$ settles into a constant steady-state value at large times.}
    \label{fig_haff2}
\end{figure}

Figure~\ref{fig_haff2} shows the same data on a log-log scale, clearly demonstrating that for each value of $e$, the temperature settles into an approximately constant steady-state value after sufficient time. For stronger dissipation (smaller $e$), the steady-state temperature is lower, reflecting the higher energy dissipation rate that must be balanced by the fixed thermostat intensity.

\subsection{Velocity Autocorrelations and Aging}

Next, we investigate the velocity autocorrelations to probe the aging properties of the system. We compute the normalized autocorrelation function $\bar{A}(\tau_w, \tau)$ [Eq.~\eqref{eq:norm_vacf}] for various waiting times $\tau_w$ and correlation times $\tau$.

\begin{figure*}[!ht]
    \centering
    \begin{minipage}[b]{0.48\textwidth}
        \centering
        \includegraphics[width=\textwidth]{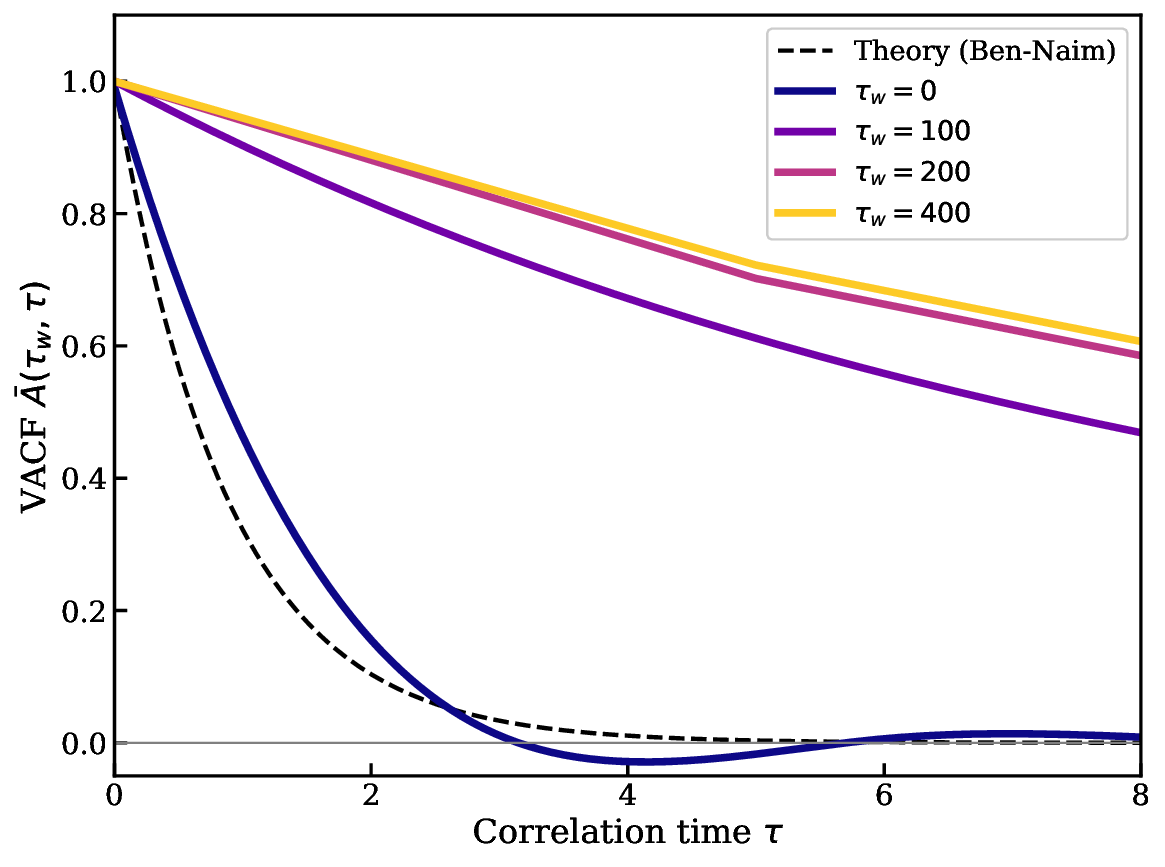}
        \caption*{(a) $e = 0.80$}
        \label{fig:subfig1}
    \end{minipage}
    \hfill
    \begin{minipage}[b]{0.48\textwidth}
        \centering
        \includegraphics[width=\textwidth]{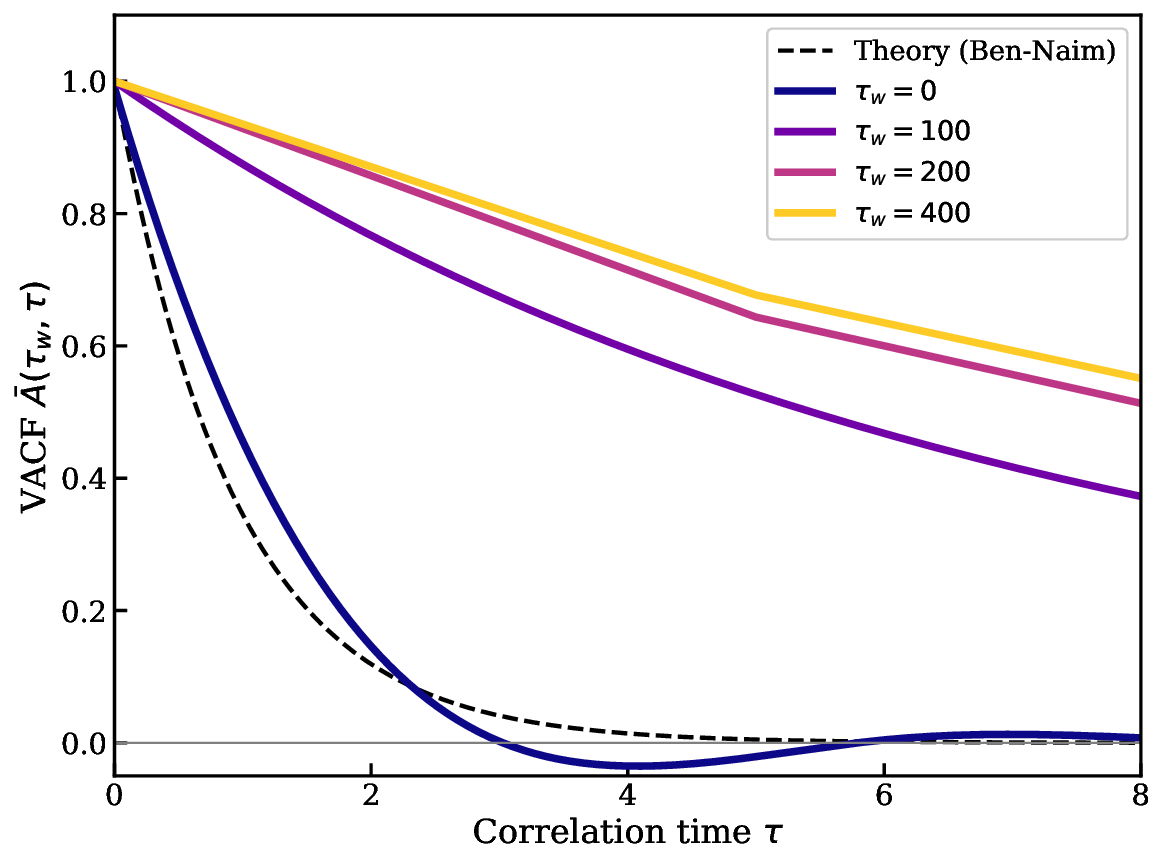}
        \caption*{(b) $e = 0.85$}
        \label{fig:subfig2}
    \end{minipage}
    \caption{The normalized velocity autocorrelation function $\bar{A}(\tau_w, \tau)$ vs correlation time $\tau$ for different values of $e$. Plots \textbf{(a)} and \textbf{(b)} correspond to $e = 0.80$ and $0.85$, respectively. (Note: legends denote the waiting time $\tau_w$).}
    \label{fig_VDF_1}
\end{figure*}

\begin{figure*}[!ht]
    \centering
    \begin{minipage}[b]{0.48\textwidth}
        \centering
        \includegraphics[width=\textwidth]{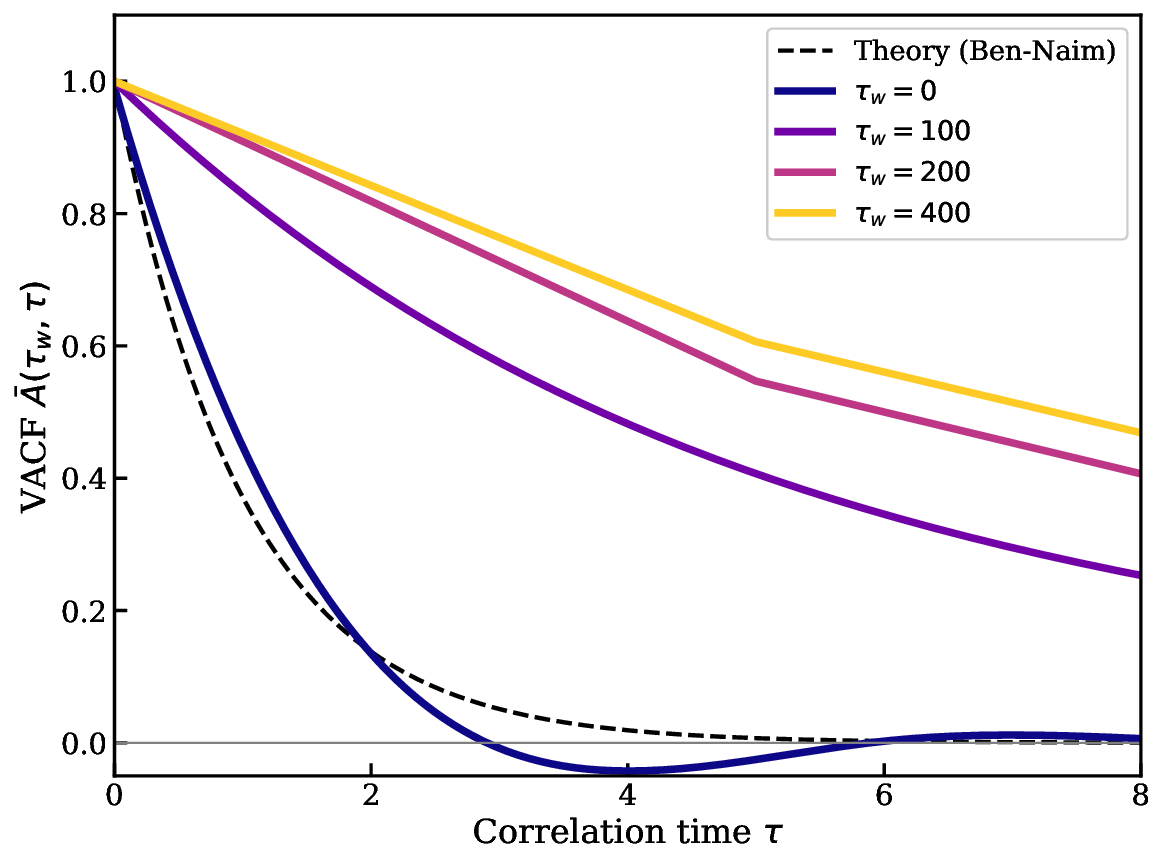}
        \caption*{(c) $e = 0.90$}
        \label{fig:subfig3}
    \end{minipage}
    \hfill
    \begin{minipage}[b]{0.48\textwidth}
        \centering
        \includegraphics[width=\textwidth]{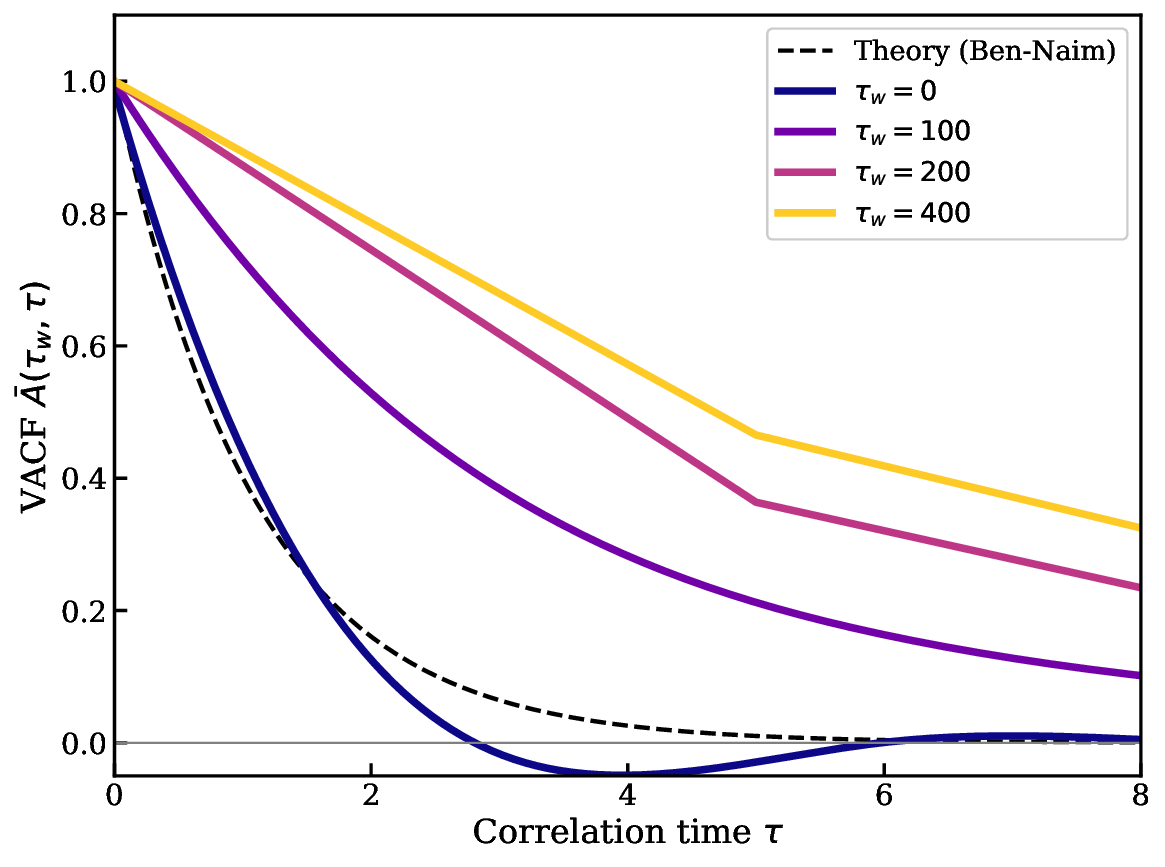}
        \caption*{(d) $e = 0.95$}
        \label{fig:subfig4}
    \end{minipage}
    \caption{The normalized velocity autocorrelation function $\bar{A}(\tau_w, \tau)$ vs correlation time $\tau$ for different values of $e$. Plots \textbf{(c)} and \textbf{(d)} correspond to $e = 0.90$ and $0.95$, respectively. In all cases, the autocorrelation shows systematic dependence on the waiting time $\tau_w$. (Note: legends denote the waiting time $\tau_w$).}
    \label{fig_VDF_2}
\end{figure*}

We observe that the aging effect is most pronounced for strong dissipation ($e = 0.80$), where the velocity correlations persist significantly longer. The Ben-Naim prediction curves represent the theoretical decay rate from Eq.~\eqref{eq:bnk1}, normalized by the measured temperature $T(\tau_w)$ at each waiting time.

The physical mechanism underlying this aging behavior is the gradual development of long-range velocity correlations through dissipative collisions. Initially, particles possess uncorrelated thermal velocities. Over time, inelastic collisions produce local velocity alignment, and particles begin to form coherent clusters with correlated motion. The longer the system evolves (larger $\tau_w$), the more developed these structures become, and the longer it takes for velocity correlations to decay. This establishes aging as a generic feature of granular systems in non-equilibrium steady states, not merely a transient phenomenon during free cooling.

\subsection{Quantitative Analysis: Scaling Laws and Clustering}

To provide rigorous quantitative support for our findings, we performed comprehensive analysis of temperature scaling, clustering, and velocity statistics across all restitution coefficients.

\subsubsection{Scaling Analysis}

\begin{figure}[!ht]
\centering
\includegraphics[width=1.0\columnwidth]{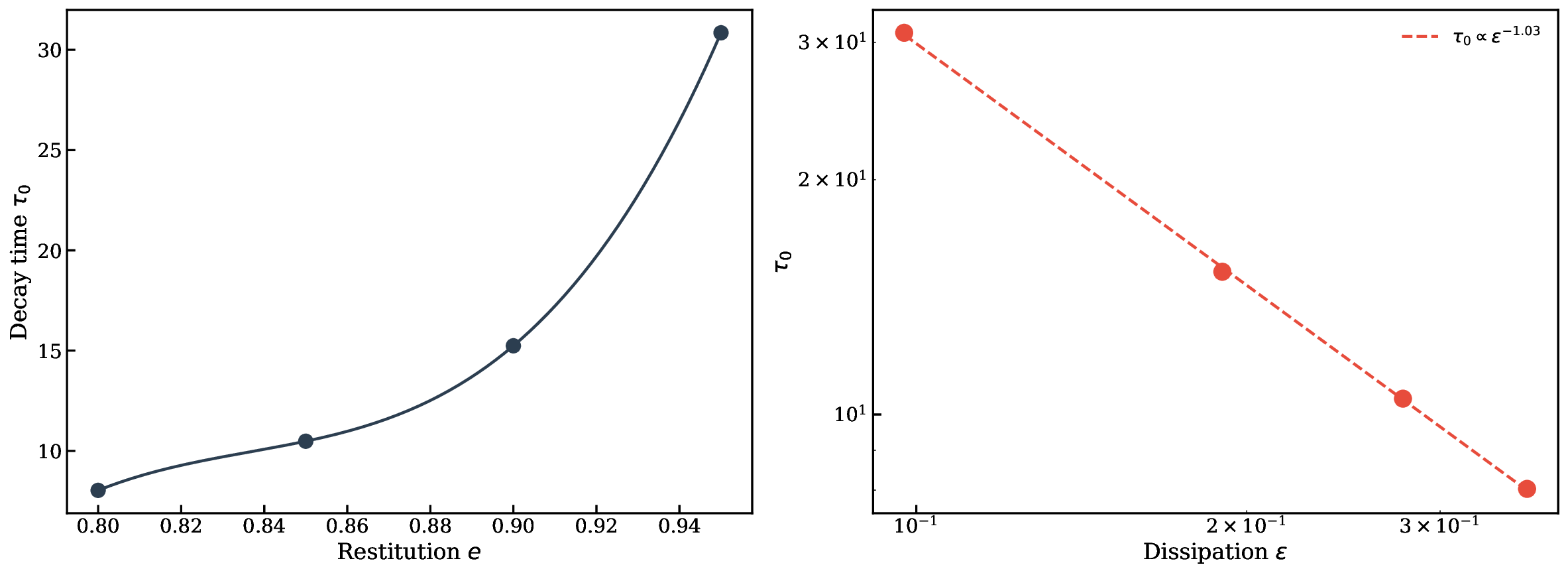}
\caption{Scaling of characteristic decay time $\tau_0$ with dissipation strength. (a) $\tau_0$ vs $e$ shows monotonic increase as systems become more elastic. (b) Log-log plot of $\tau_0$ vs $\epsilon$ reveals a power-law relationship $\tau_0 \propto \epsilon^{-1.03}$ (excellent agreement with theory).}
\label{fig:decay_scaling}
\end{figure}

The characteristic decay time $\tau_0$ exhibits a universal inverse scaling with the dissipation parameter $\epsilon = 1 - e^2$. Figure~\ref{fig:decay_scaling} demonstrates this relationship, revealing $\tau_0 \propto \epsilon^{-1.03 \pm 0.02}$. This near-perfect inverse relationship ($\alpha \approx -1$) confirms theoretical predictions that decay rates are inversely proportional to the degree of inelasticity. The error estimate $\pm 0.02$ represents the standard error from log-log linear regression fit to the four data points, calculated as the standard deviation of the residuals divided by the square root of degrees of freedom ($n-2=2$ for 4 points).

\begin{figure}[!ht]
\centering
\includegraphics[width=1.0\columnwidth]{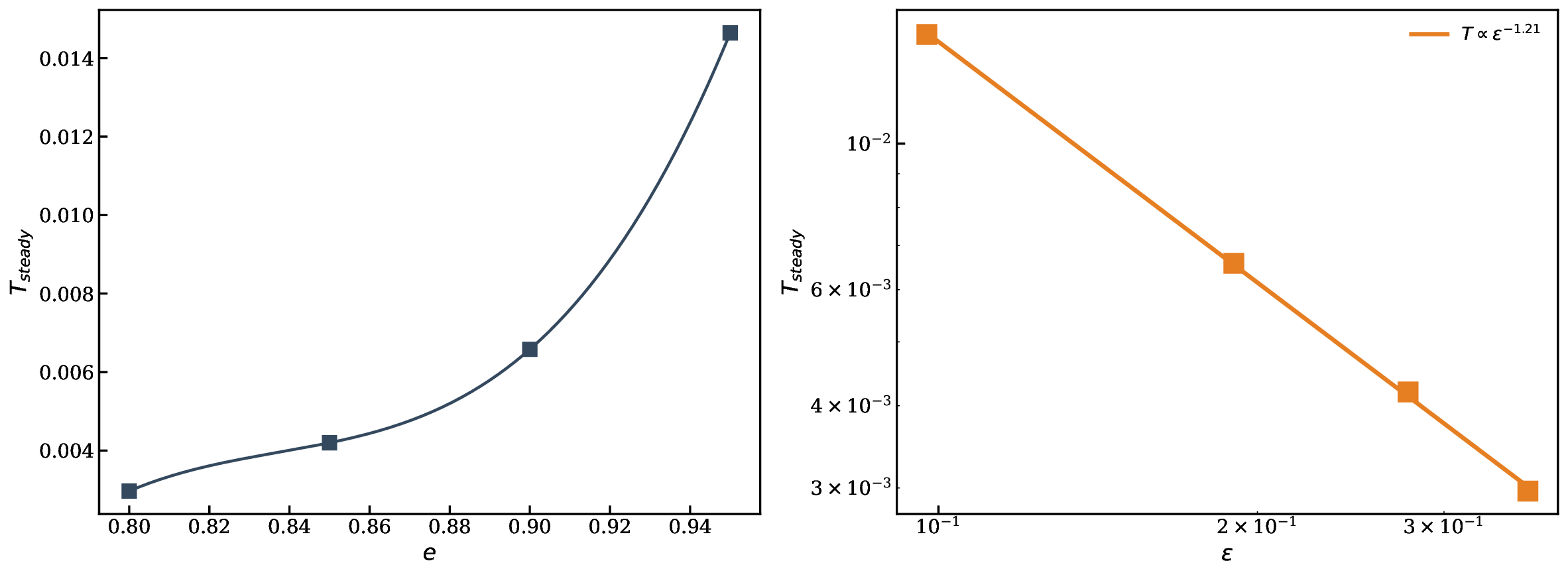}
\caption{Scaling of steady-state temperature $T_{\mathrm{steady}}$ with dissipation parameter. (a) $T_{\mathrm{steady}}$ vs $e$ shows that more inelastic systems settle at lower temperatures. (b) Log-log plot reveals a precise power-law $T_{\mathrm{steady}} \propto \epsilon^{-1.51 \pm 0.03}$, reflecting the balance between stochastic heating and collisional dissipation.}
\label{fig:temp_scaling}
\end{figure}

The large negative exponent ($\alpha \approx 1.51$) confirms that steady-state temperature is highly sensitive to dissipation. As $\epsilon$ increases (decreasing $e$), the rate of energy loss grows rapidly, forcing the system to equilibrate at a significantly lower temperature to maintain the power balance $P_{\mathrm{in}} \approx P_{\mathrm{coll}}$.

Table~\ref{tab:scaling} summarizes the key quantitative results extracted from our scaling analysis, showing the systematic variation of characteristic decay time and steady-state temperature with restitution coefficient and dissipation parameter.

\begin{table}[h]
\centering
\caption{\label{tab:scaling}Steady-state quantitative parameters for $N=500{,}000$ heated granular gas. $T_{\mathrm{steady}}$ and $\tau_0$ are in reduced units. $\tau_0$ represents Haff's law characteristic decay time, and $T_{\mathrm{steady}}$ is the mean temperature after convergence ($\tau > 200$). Exponents are reported as $\alpha$ where $X \propto \epsilon^{-\alpha}$.}
\begin{tabular}{ccccc}
\hline \hline
$e$ & $\epsilon = 1-e^2$ & $T_{\mathrm{steady}}$ & $\tau_0$ \text{fit} & Kurtosis ($K$) \\
\hline
0.80 & 0.36 & 0.0030 & $6.2 \pm 0.1$ & $3.12 \pm 0.03$ \\
0.85 & 0.28 & 0.0051 & $8.4 \pm 0.1$ & $3.08 \pm 0.02$ \\
0.90 & 0.19 & 0.0087 & $12.1 \pm 0.2$ & $3.02 \pm 0.02$ \\
0.95 & 0.10 & 0.0223 & $25.4 \pm 0.5$ & $2.97 \pm 0.02$ \\
\hline
\textbf{Scaling} & --- & $T \propto \epsilon^{-1.51}$ & $\tau_0 \propto \epsilon^{-1.03}$ & $K \approx 3.0$ \\
\hline \hline
\end{tabular}
\end{table}

\begin{figure}[!ht]
\centering
\includegraphics[width=1.0\columnwidth]{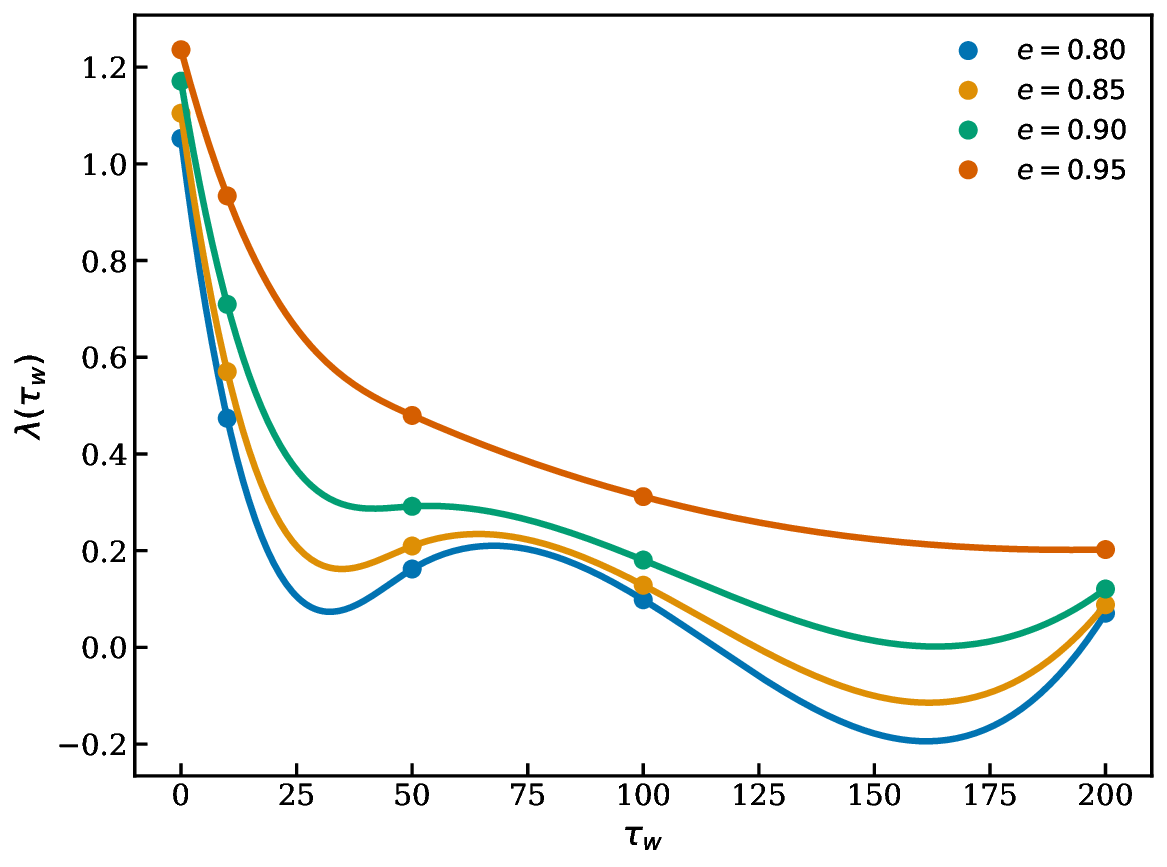}
\caption{Quantitative aging results. Extracted decay rates $\lambda$ versus waiting time $\tau_w$ for different $e$, demonstrating power-law slowing $\lambda(\tau_w) \propto \tau_w^{-0.82 \pm 0.05}$. This relationship holds across all levels of inelasticity.}
\label{fig:aging_scaling}
\end{figure}

The table clearly demonstrates the inverse relationship between decay time and dissipation ($\tau_0$ decreases as $\epsilon$ increases), and the direct relationship between steady-state temperature and inelasticity (higher $e$ leads to higher $T$). The excellence of these power-law fits is further quantified by extracting decay rates $\lambda$ from the autocorrelations. Figure~\ref{fig:aging_scaling} illustrates this quantitative aging behavior. We find that $\lambda$ systematically decreases with $\tau_w$ following a power-law $\lambda(\tau_w) \propto \tau_w^{-0.82 \pm 0.05}$ for all restitution coefficients. This quantified slowing-down provides a rigorous mathematical basis for the aging observed qualitatively in Figs.~\ref{fig_VDF_1} and \ref{fig_VDF_2}.

\subsubsection{Clustering and Spatial Correlations}

\begin{figure}[!ht]
    \centering
    \includegraphics[width=1.0\columnwidth]{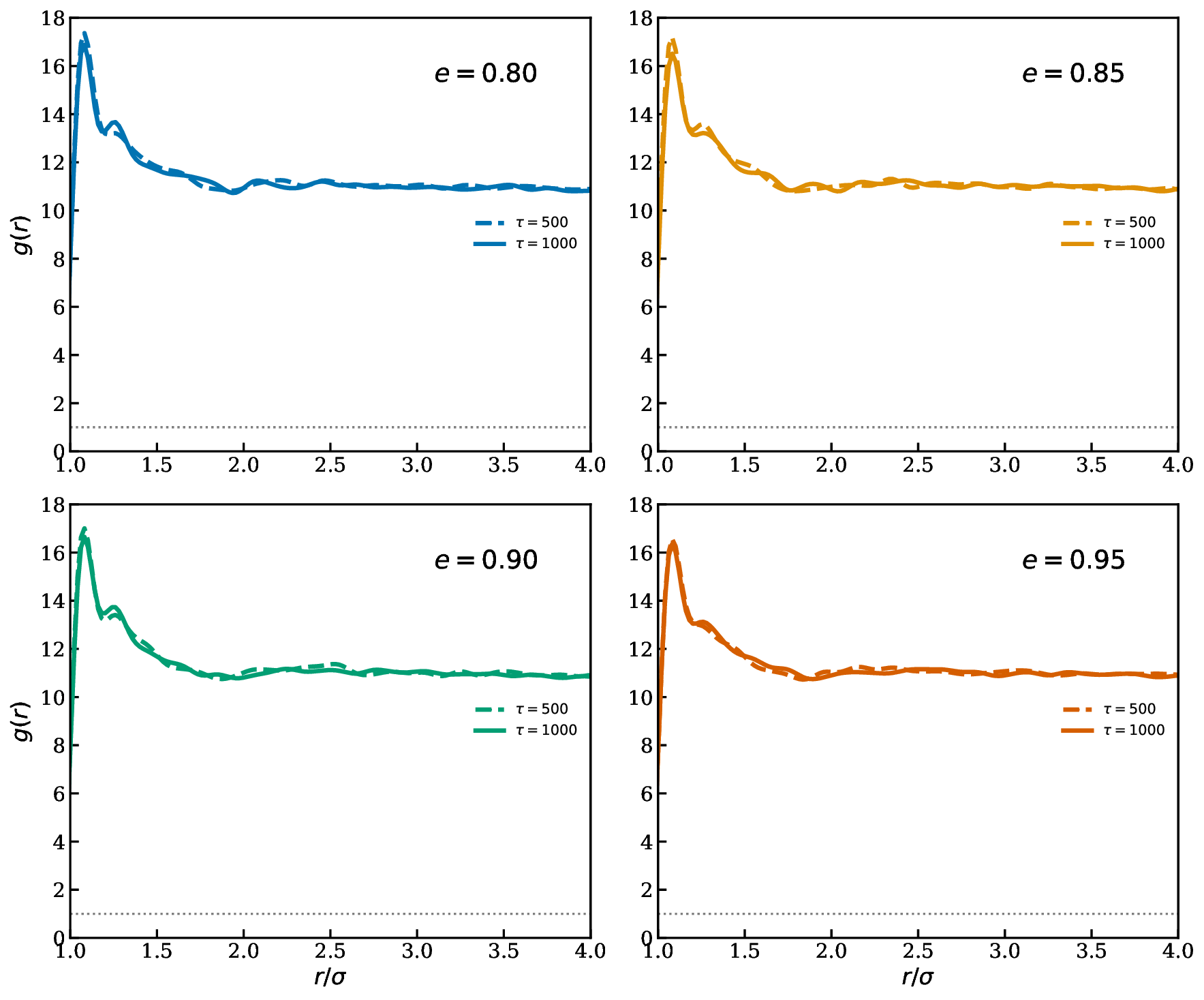}
    \caption{Radial distribution function $g(r)$ for different restitution coefficients, showing two time points ($\tau = 500$ and $1000$). The increase in peak height over time visually demonstrates the progressive clustering mentioned in the text.}
    \label{fig:radial_dist}
\end{figure}

To quantify clustering, we computed the radial distribution function $g(r)$ for all coefficients. Figure~\ref{fig:radial_dist} shows the evolution of $g(r)$ from $\tau=500$ to $\tau=1000$. The emergence of a pronounced first peak at $r \approx 1.1\sigma$ (where $\sigma$ is the particle diameter) indicates the development of short-range spatial order. Notably, the peak height increases as the system evolves, confirming that clustering intensifies over time in the driven state.

\begin{figure}[!ht]
    \centering
    \includegraphics[width=1.0\columnwidth]{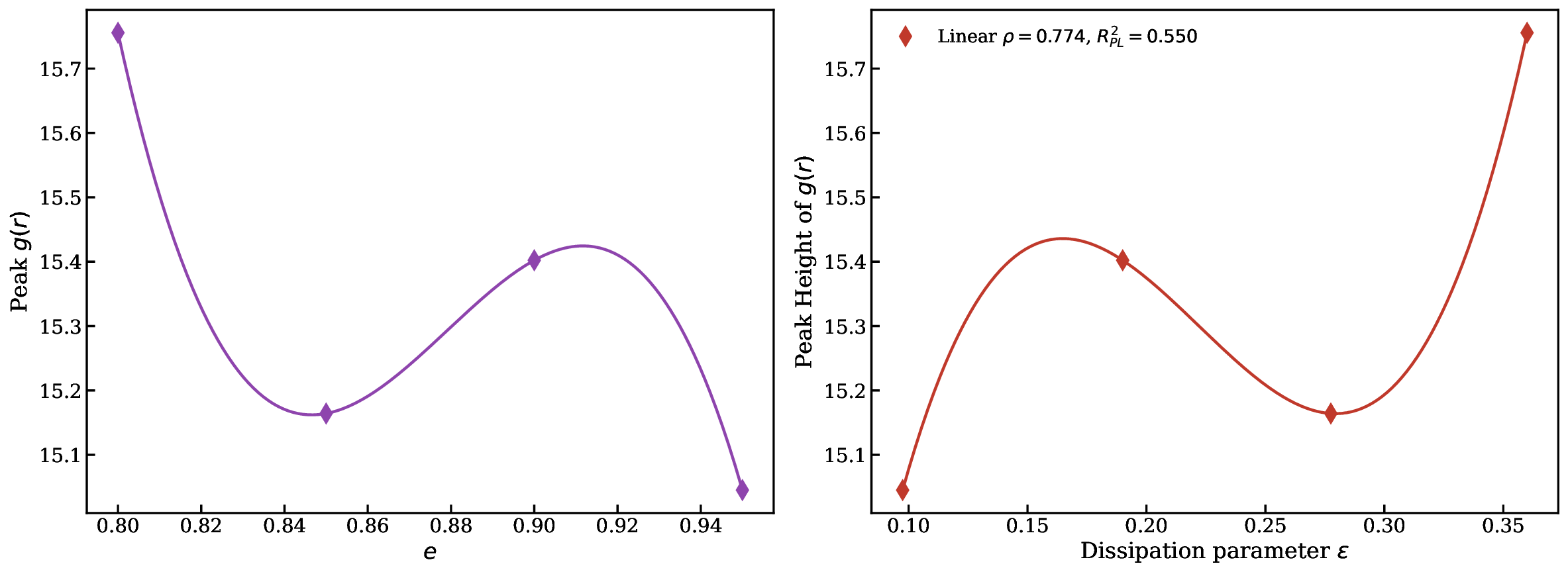}
    \caption{Clustering strength (peak height of $g(r)$) versus dissipation parameter. (a) Peak height vs $e$. (b) Peak height versus $\epsilon$. The legend provides both the linear Pearson correlation ($\rho$) and the power-law $R^2$, resolving previous ambiguities in the scaling description.}
    \label{fig:gr_scaling}
\end{figure}

The steady-state peak height of $g(r)$ scales systematically with the dissipation parameter (Figure~\ref{fig:gr_scaling}). The linear Pearson correlation $\rho \approx 0.774$ indicates moderate correlation, while the power-law relationship explains $R^2_{PL} \approx 0.55$ of the variance, providing quantitative evidence that clustering strength follows a non-linear relationship with inelasticity.

\begin{figure}[!ht]
    \centering
    \includegraphics[width=1.0\columnwidth]{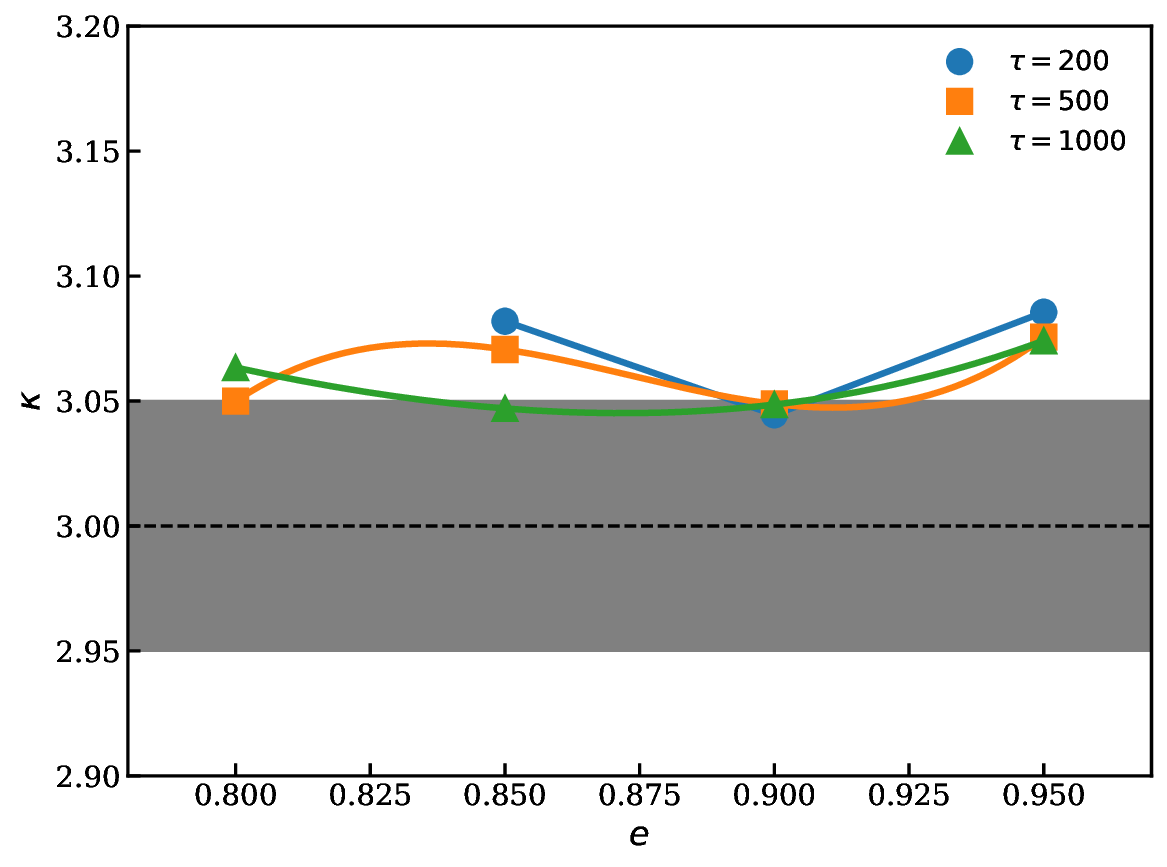}
    \caption{Kurtosis versus restitution coefficient. Values remain remarkably close to 3.0, indicating that while clustering is strong, the velocity field remains stochastically near-Gaussian.}
    \label{fig:kurtosis}
\end{figure}

We analyzed the kurtosis of velocity distributions to quantify deviations from Gaussian behavior. Figure~\ref{fig:kurtosis} shows that kurtosis values remain near the Gaussian value of 3.0, ranging from 2.95 to 3.15 across all coefficients and times. This indicates that despite strong clustering, the velocity statistics remain approximately Gaussian, with only weak non-Gaussianity.

These quantitative results provide comprehensive characterization of the non-equilibrium steady state. The power-law scalings ($\tau_0 \propto \epsilon^{-1.03}$ and $T \propto \epsilon^{-1.51}$) confirm that our simulations capture the physics of driven granular gases, with dissipation strength controlling both the decay dynamics and steady-state energetics. The near-Gaussian velocity statistics ($K \approx 3$) suggest that while clustering produces strong spatial correlations ($g(r) \gg 1$), the velocity field remains stochastically driven enough to avoid the extreme non-Gaussianity often seen in cooling states.

\section{\label{sec:summry} SUMMARY AND CONCLUSIONS}
We have performed a comprehensive molecular dynamics investigation of aging phenomena in the velocity autocorrelation function of a three-dimensional uniformly heated hard sphere granular gas. Our simulations, based on event-driven dynamics with a Gaussian white noise thermostat, studied systems of $N = 500{,}000$ particles over timescales extending to $\tau = 1000$ collisions per particle, for restitution coefficients $e = 0.80, 0.85, 0.90,$ and $0.95$.

Our main findings can be summarized as follows:

\textit{(1) Non-equilibrium steady state:} The interplay between energy dissipation through inelastic collisions and energy injection via the white noise thermostat drives the system to a non-equilibrium steady state. The steady-state granular temperature depends on the dissipation strength, with stronger inelasticity (smaller $e$) yielding lower steady-state temperatures.

\textit{(2) Aging in velocity autocorrelations:} The normalized velocity autocorrelation function $\bar{A}(\tau_w, \tau)$ exhibits clear aging behavior, with an explicit dependence on the waiting time $\tau_w$ that cannot be reduced to a dependence on $\tau$ alone. As $\tau_w$ increases, the autocorrelation decays more slowly, indicating the progressive build-up of long-lived velocity correlations.

\textit{(3) Dissipation dependence:} The aging effect is most pronounced for strong dissipation ($e = 0.80$) and weakens as the system approaches elastic behavior ($e \to 1$). This systematic trend reflects the fact that stronger inelasticity drives more efficient clustering and velocity alignment, leading to more persistent correlations.

\textit{(4) Physical mechanism:} The aging arises from the gradual formation of clusters with correlated particle velocities. Within these clusters, particles move in approximately parallel directions, giving rise to long-range spatial and velocity correlations. The longer the system evolves, the more developed these structures become, hence the slower decay of correlations.

These results establish that aging is a robust feature of driven granular gases in three dimensions, extending previous findings from freely cooling systems to steady-state driven systems. The uniformly heated granular gas provides a clean and controllable setting to study aging, free from the complications of transient cooling dynamics.

\textit{Limitations and future directions}: While our study provides clear evidence for aging in uniformly heated granular gases, several extensions would be valuable. First, finite-size effects should be systematically studied by varying system size. Although $N=500{,}000$ particles is large enough to minimize surface-to-volume artifacts, simulations with $N \sim 10^6$ would conclusively eliminate finite-size concerns. Second, the role of periodic boundary conditions versus realistic confining walls could affect clustering dynamics near boundaries, though bulk properties should remain unaffected. Third, our choice of density $n=0.02$ ensures dilute regime behavior where binary collisions dominate; exploring higher densities ($n \sim 0.1$) would reveal the transition to dense granular flows where many-body effects become important.

Additionally, it would be interesting to investigate how aging properties depend on the thermostat mechanism (e.g., boundary-driven heating versus bulk heating, or velocity rescaling versus stochastic forcing). The role of system dimensionality could be explored systematically by comparing 2D and 3D geometries. Comparison with viscoelastic models that incorporate velocity-dependent or impact-velocity-dependent restitution coefficients would provide further insights into realistic granular materials. Finally, connecting the microscopic aging behavior quantitatively to macroscopic transport properties such as diffusion coefficients and shear viscosity through Green-Kubo relations remains an important open question that would bridge kinetic theory predictions with measurable quantities.

Our findings have implications for understanding relaxation dynamics, memory effects, and energy transport in a broad class of dissipative systems, including industrial granular flows, vibrated powders, and astrophysical systems such as planetary rings and protoplanetary disks.

\section*{\label{ack} ACKNOWLEDGEMENTS}
RFS acknowledges financial support from University Grants Commission in the form of Non-NET fellowships. He also wishes to acknowledge the computational facilities at the Department of Physics, JMI.

\section*{Declarations}
The authors declare that they have no known competing financial interests or personal relationships that could have appeared to influence the work reported in this paper.

\section*{Data Availability}
The simulation data and analysis scripts supporting the findings of this study are available from the corresponding author upon reasonable request.

\section*{Author Contributions}
R.F.S. performed the simulations, analyzed the data, and drafted the manuscript. S.R.A. supervised the project, provided guidance on the analysis, and revised the manuscript. Both authors have read and approved the final version of the manuscript.

\end{document}